\documentclass[12pt]{article}

\usepackage{amsfonts}
  \usepackage{amsthm}
 \usepackage{amsmath}
 \usepackage{amscd}
 \usepackage{amssymb}
 \usepackage{latexsym}

 \numberwithin{equation}{section}
 \newtheorem{thm}{Theorem}
 
 \newtheorem{prop}{Proposition}
 \theoremstyle{definition}

\title{Near-integrability of low dimensional periodic Klein-Gordon lattices}
\author{Ognyan Christov \\
Faculty of Mathematics and Informatics, Sofia University, \\
5 J. Bourchier blvd. 1164 Sofia, Bulgaria}
\date{}

\begin{document}

\maketitle

\begin{abstract}
\noindent The low dimensional periodic Klein-Gordon lattices are
studied for integrability. We prove that the periodic lattice with
two particles and certain nonlinear potential is non integrable.
However, in the cases of up to six particles, we prove that their
Birkhoff-Gustavson normal forms are integrable, which allows us to
apply KAM theory.
\end{abstract}

\section{Introduction}

In this article we deal with the  periodic Klein-Gordon (KG)
lattice (see for example \cite{Morgante} and references therein)
described by the Hamiltonian
\begin{equation}
\label{1.1} H = \sum_{j \in \mathbb{Z}/n\mathbb{Z}}
\Big[\frac{p_j^2}{2} + \frac{C}{2}(q_{j+1} - q_j)^2 + V (q_j)
\Big], \quad p_j = \dot{q}_j .
\end{equation}
The constant $C > 0$ measures the interaction to nearest neighbor
particles (with unit masses) and $V (x)$ is a non-linear
potential.

We study the integrability of (\ref{1.1}). When $C=0$ the
Hamiltonian is separable and, hence integrable. There exist plenty
of periodic or quasi-periodic solutions in the dynamics of
(\ref{1.1}). It is natural to investigate whether this behavior
persists for $C$ small enough (see e.g. \cite{Jong}). Here we do
not assume that $C$ is small.

We are interested in the behavior at low energy, that is why the
following  main assumptions are in order:

\begin{itemize}
\item  $V (x) = \frac{a}{2} x^2 + \frac{b}{2} x^4$,
\item  $a > 0 $ irrational.
\end{itemize}
{\bf Remark 1.} Such type of potentials are frequently used in the
literature \cite{Morgante}). As it can be seen below the choice of
$a$ simplifies considerably the calculations.

\vspace{2ex}

We can also assume that $C=1$ which can be achieved by rescaling
of $t$. Then our Hamiltonian takes the form
\begin{equation}
\label{1.2} H = \sum_{j \in \mathbb{Z}/n\mathbb{Z}}
\Big[\frac{p_j^2}{2} + \frac{1}{2}(q_{j+1} - q_j)^2 +
\frac{a}{2}(q_j)^2 + \frac{b}{2} (q_j)^4 \Big], \quad p_j =
\dot{q}_j .
\end{equation}
Our first result concerns the Hamiltonian with two degrees of
freedom ($q_2 = q_0$), i.e.
\begin{equation}
\label{a1} H = \frac{1}{2}\left(p_1^2 + p_2^2\right) +
\frac{1}{2}\left(2 q_1^2 - 4 q_1 q_2 + 2 q_2^2\right) +
\frac{a}{2}\left(q_1^2 + q_2^2\right) + \frac{b}{2}\left(q_1^4 +
q_2^4\right).
\end{equation}
It simply says that the corresponding Hamiltonian system is
integrable only when it is linear.
\begin{thm}
\label{th1} The periodic KG lattice with $n=2$ is non-integrable
unless $b=0$.
\end{thm}
The above result tells us that it is highly unlikely to expect
integrability for $n > 2$ (see also the discussion in the end of
the paper).

\vspace{3ex}

Motivated by the works of Rink \cite{RinkVer,Rink2}, who presented
the periodic FPU chain as a perturbation of an integrable and KAM
non-degenerated system, namely the truncated Birkhoff-Gustavson
normal form of order 4 in the neighborhood of an equilibrium, our
aim is to verify whether this can be done for the low dimensional
KG lattices.

One should note that the Rink's result is due to the special
symmetry and resonance properties of the FPU chain and should not
be expected for lower-order resonant Hamiltonian systems (see e.g.
\cite{Ch1}).

We summarize  our second result in the following
\begin{thm}
\label{th2}
The truncated normal forms $\overline{H} = H_2 +
\overline{H}_4 $ of the periodic KG lattices up to six particles
 are completely integrable. In particular, these normal forms are KAM
 non-degenerated excepting the case of six particles.
\end{thm}
As a consequence from this result, we may conclude for the low
dimensional KG lattices when KAM theory applies, that there exist
many quasi-periodic solutions of small energy on a long time scale
(see section 2 and for more detailed explanation \cite{RinkVer})
and chaotic orbits are of small measure.

\vspace{3ex}

The paper is organized as follows. In section 2 some notions and
facts used in the paper are given. In section 3 we calculate the
Birkhoff-Gustavson normal forms for the cases up to six particles
and show that they are integrable. We finish with some concluding
remarks as well as some possible lines of further study.

The proof of Theorem \ref{th1} is based on the
Ziglin-Morales-Ramis theory and since it is more algebraic in
nature, it is carried out in the Appendix.

\section{Resonances and normalization}

In this section we recall briefly some notions and facts about
integrability of Hamiltonian systems, action-angle variables,
perturbation of integrable systems and normal forms. More complete
exposition can be found in \cite{A1,AbrMars,AKN}.

 Let $H$ be an analytic Hamiltonian defined on a $2 n$ dimensional
symplectic manifold. The corresponding
 Hamiltonian system is
\begin{equation}
\label{2.1}
 \dot x = X_H (x).
 \end{equation}
It is said that a Hamiltonian system is completely integrable if there exist $n$
independent integrals $F_1 = H, F_2, \ldots, F_n$ in involution,
namely $\{ F_i, F_j \} = 0$ for all $i$ and $j$, where $\{ , \}$
is the Poisson bracket. On a neighborhood $U$ of the connected
compact level sets of the integrals $M_c = \{ F_j = c_j, j = 1,
\ldots, n \}$  by Liouville - Arnold theorem one can introduce a
special set of symplectic coordinates, $ I_j, \varphi_j$, called
action - angle variables. Then, the integrals $F_1 = H, F_2,
\ldots, F_n$  are  functions of action variables only and the flow
of $X_H$ is simple
\begin{equation}
\label{2.2}
\dot{I}_j = 0, \quad \dot{\varphi}_j = \frac{\partial
H}{\partial I_j}, \quad j = 1, \ldots, n.
\end{equation}
Therefore, near $M_c$, the phase space is foliated with $X_{F_i}$
invariant tori over which the flow of $X_H$ is quasi - periodic
with frequencies $ (\omega_1 (I), \ldots, \omega_n (I) ) = (
\frac{\partial H}{\partial I_1}, \ldots, \frac{\partial
H}{\partial I_n})$.

The map
\begin{equation}
\label{2.3} (I_1, I_2, \ldots, I_n) \to \left( \frac{\partial
H}{\partial I_1}, \frac{\partial H}{\partial I_2}, \ldots,
\frac{\partial H}{\partial I_n}\right)
\end{equation}
is called frequency map.

Consider a small perturbation of an integrable Hamiltonian $H_0$.
According to Poincar\'{e} the main problem of mechanics is to
study the perturbation of quasi-periodic motions in the system
given by the Hamiltonian
$$
H = H_0 (I) + \varepsilon H_1 (I, \varphi), \quad \varepsilon < <
1.
$$
 KAM - theory \cite{K,A,Moser} gives conditions on the integrable Hamiltonian $H_0$ which ensures
the survival of the most of the invariant tori. The following
condition, usually called Kolmogorov's condition, is that the
frequency map should be a local diffeomorphism, or equivalently
\begin{equation}
\label{2.4} \det\left( \frac{\partial^2 H_0}{\partial I_i \partial
I_j}  \right) \neq 0
\end{equation}
on an open and dense  subset of $U$. We should note that the
measure of the surviving tori decreases with the increase of both
perturbation and the measure of the set where above Hessian is too
close to zero.


 In the neighborhood of an equilibrium $(0, 0)$ we have the
 following expansion of $H$
\begin{align*}
H   & = H_2 + H_3 + H_4 + \ldots, \\
H_2 & = \sum \omega_j ( q_j ^2 + p_j ^2 ), \quad \omega_j > 0 .
\end{align*}
We assume that $H_2$ is a positively defined quadratic form. The
frequency $\omega = (\omega_1, \ldots, \omega_n)$ is said to be in
resonance if there exists a vector $k = (k_1, \ldots, k_n), \,
k_j \in \mathbb{Z}, j = 1, \ldots, n$, such that $(\omega, k) =
\sum k_j \omega_j = 0$,  where $|\, k | = \sum |\, k_j |$  is the
order of resonance.

 With the help of a series of canonical transformations
close to the identity, $H$  simplifies. In the absence of
resonances the simplified Hamiltonian is called  Birkhoff normal
form, otherwise -  Birkhoff-Gustavson normal form.

Often to detect the behavior in a small neighborhood of the
equilibrium, instead of the Hamiltonian $H$ one considers the
normal form truncated to some order
$$
\overline{H} = H_2 + \ldots + \overline{H}_m .
$$
It is known that the truncated to any order Birkhoff normal form
 is integrable \cite{AKN}. The truncated Birkhoff-Gustavson normal
 form has at least two integrals - $H_2$ and $\bar{H}$.  Therefore, the truncated
normal form of two degrees of freedom Hamiltonian is integrable.

In order to obtain estimates of the approximation by normalization
in a neighborhood of an equilibrium point we scale $q \to
\varepsilon \tilde{q}, \, p \to \varepsilon \tilde{p}.$ Here
$\varepsilon$ is a small positive parameter and $\varepsilon^2$ is
a measure for the energy relative to the equilibrium energy. Then,
dividing by $\varepsilon^2$ and removing tildes we get
$$
 \overline{H} = H_2 + \varepsilon \overline{H}_3 + \ldots + \varepsilon^{m-2} \overline{H}_m .
$$
 Provided that $\omega_j > 0$ it is proven  in \cite{V1} that $\bar{H}$
is an integral for the original system with error $O
(\varepsilon^{m-1})$
 and $H_2$ is an integral for the original system with error $O (\varepsilon)$ for  the whole time interval.
 If we have more  independent integrals, then they are integrals
 for the original Hamiltonian system with error $O (\varepsilon^{m-2})$
 on the time scale $1/\varepsilon$.

 The first integrals for the normal form  $\overline{H}$ are approximate integrals for the original system,
 that is,   if the normal form is integrable then the original system is
 {\it near integrable}  in the above sense.


Returning to the Hamiltonian of the periodic KG lattice
(\ref{1.2}) we see that its
 quadratic part $H_2$  is not in diagonal form
\begin{equation}
\label{2.5} H_2 = \frac{1}{2} p^T p + \frac{1}{2} q^T L_n q.
\end{equation}
Here $L_n$ is the following $n \times n$ matrix
\begin{equation}
\label{2.6} L_n :=
\begin{pmatrix}
2+a & -1 &   &   & -1 \\
-1 & 2+a & -1  &   &  \\
\\
   & \ddots & \ddots   & \ddots       &  \\
   \\
  &   &-1   &  2+a & -1   \\
-1 &   &   & -1 & 2+a
\end{pmatrix} .
\end{equation}
The eigenvalues of $L_n$ are of the form $\Omega_k = a + \omega^2
_k, \, \omega_k = 2 \sin \frac{k \pi}{n}$. In order to obtain the
corresponding eigenvectors $y^k$, following \cite{RinkVer} we
define
\begin{equation}
\label{2.7} y^n := \frac{1}{\sqrt{n}} (1, 1, \ldots, 1)^T
\end{equation}
and if $n$ is even,
\begin{equation}
\label{2.8} y^{n/2} := \frac{1}{\sqrt{n}} (1, -1, 1, -1, \ldots,
-1)^T .
\end{equation}
Further, for $1 \leq k < n/2$, we define $y^k$ and $y^{n-k}$ via
their coordinates
\begin{equation}
\label{2.9} y^k _j := \sqrt{\frac{2}{n}} \cos \left( \frac{2 k j
\pi}{n} \right) , \quad y^{n-k} _j := \sqrt{\frac{2}{n}} \sin
\left( \frac{2 k j \pi}{n} \right) .
\end{equation}
It is easily checked that $\{ y^1, \ldots, y^n \}$ is an
orthonormal basis of $\mathbb{R}^n$, consisting of eigenvectors of
$L_n$. Let $Y$ be the $n \times n $ matrix formed by the vectors
$y^k$ as columns, then $Y^T Y = Id, \, Y^{-1} L_n Y = \Omega :=
\mathrm{diag} (\Omega_1, \ldots, \Omega_n)$. The symplectic
Fourier-transformation $q = Y \bar{q}, p = Y \bar{p}$ brings $H_2$
in diagonal form
\begin{equation}
\label{2.10} H_2 = \frac{1}{2} p^T p + \frac{1}{2} q^T \Omega q .
\end{equation}
The variables $(\bar{q}, \bar{p})$ are known as {\it phonons}.

We need one more definition.

{\bf Definition.} (\cite{RinkVer}) It is said that $\omega \in
\mathbb{R}^n$ satisfies the property of {\it internal resonance}
if for any $k \in \mathbb{Z}^n$ with $(k, \omega) = 0$, \, $ | k |
= 4$, we have $k_j = -k_{n-j}$ when $1 \leq j < n/2$.

In the following table we list the frequencies $\Omega_k$ of some
low dimensional periodic KG lattice:

\begin{table}[ht]
{\begin{tabular}{c|l}
n & $\Omega_k$  \\[1ex]
 \hline
  $2$ & $\sqrt{a+4}$, \, $\sqrt{a}$ \\[1ex]
 $3$ & $\sqrt{a+3}$, \, $\sqrt{a+3}$, \, $\sqrt{a}$ \\[1ex]
 $4$ & $\sqrt{a+2}$, \, $\sqrt{a+4}$, \, $\sqrt{a+2}$, \, $\sqrt{a}$ \\[1ex]
 $5$ & $\sqrt{a+(5-\sqrt{5})/2}$, \, $\sqrt{a+(5+\sqrt{5})/2}$ , \,  $\sqrt{a+(5+\sqrt{5})/2}$, \,  $\sqrt{a+(5-\sqrt{5})/2}$, \, $\sqrt{a}$ \\[1ex]
 $6$ & $\sqrt{a+1}$, \, $\sqrt{a+3}$, \, $\sqrt{a+4}$, \,  $\sqrt{a+3}$, \, $\sqrt{a+1}$, \, $\sqrt{a}$
\end{tabular}}
\end{table}

As it is seen from the table we almost always have internal resonances.
The assumption on $a$ prevents the appearance of more complicated
resonances.

\section{Low-dimensional lattices}

In this section we calculate the normal forms for the periodic KG
lattices with particles up to six. We do not use $\varepsilon$ in
the forth degree expression, but keep in mind that we are close to
the equilibrium. Of course, it is assumed that $b \neq 0$.

\subsection{Two particles}
This case is easy. It is well known that the truncated to any
order normal form of a two degrees of freedom Hamiltonian is
integrable. It remains only to verify the KAM condition.

We have already brought the quadratic part of the Hamiltonian in
diagonal form. It is important that it is written in the phonons
$(\bar{q}, \bar{p})$.
\begin{equation}
\label{3.1}
H = \frac{1}{2}\left(\bar{p}_1^2 + \bar{p}_2^2\right)
+ \frac{1}{2}\big[(4+a) \bar{q}_1^2 + a \bar{q}_2^2 \big] +
\frac{b}{4}\left(\bar{q}_1^4 + 6 \bar{q}_1^2 \bar{q}_2^2 +
\bar{q}_2^4\right).
\end{equation}
Further, we perform a scaling
\begin{eqnarray*}
\bar{q}_1 & \to  \frac{1}{\sqrt[4]{a+4}} \bar{q}_1, \quad \bar{p}_1 &  \to  \sqrt[4]{a+4} \bar{p}_1 , \\
\bar{q}_2 & \to  \frac{1}{\sqrt[4]{a}} \bar{q}_2,   \quad
\bar{p}_2 &  \to  \sqrt[4]{a} \bar{p}_2 ,
\end{eqnarray*}
which preserves the symplectic form. The Hamiltonian (\ref{3.1})
becomes
$$
H = \frac{\sqrt{a+4}}{2} (\bar{p}_1^2 + \bar{q}_1^2) +
\frac{\sqrt{a}}{2} (\bar{p}_2^2 + \bar{q}_2^2) +
\frac{b}{4}\left(\frac{\bar{q}_1^4}{a+4} + \frac{6\bar{q}_1^2
\bar{q}_2^2}{\sqrt{a(a+4)}} + \frac{\bar{q}_2^4}{a} \right).
$$
Usually at this place one makes  the following change of variables
\begin{equation}
\label{3.2} \bar{q}_j = \frac{1}{2} (z_j + w_j), \quad \bar{p}_j =
\frac{1}{2 i} (z_j - w_j), \quad j = 1, 2.
\end{equation}
Since the frequencies $\Omega_1 = \sqrt{a+4}, \Omega_2 = \sqrt{a}$
are incommensurable, the only resonant terms which remain are
$$
z_j w_j, \quad (z_j w_j)^2, \, j=1,2, \quad z_1 w_1 z_2 w_2 .
$$
The other terms can be removed via symplectic near-identity
change. Therefore, the normal form of (\ref{3.1}) up to order 4
$\overline{H} = H_2 + \overline{H}_4$ is
$$
\overline{H} = \frac{\sqrt{a+4}}{2} z_1 w_1 + \frac{\sqrt{a}}{2}
z_2 w_2 + \frac{3b}{32} \Big[\frac{(z_1 w_1)^2}{a+4} + 4 \frac{z_1
w_1 z_2 w_2}{\sqrt{a(a+4)}} + \frac{(z_2 w_2)^2}{a} \Big]
$$
or, returning to the  $(\bar{q}, \bar{p})$ coordinates we get
$$
\overline{H} = \frac{\sqrt{a+4}}{2} (\bar{p}_1^2 + \bar{q}_1^2) +
\frac{\sqrt{a}}{2} (\bar{p}_2^2 + \bar{q}_2^2) + \frac{3b}{32}
\Big[\frac{(\bar{p}_1^2 + \bar{q}_1^2)^2}{a+4} + 4
\frac{(\bar{p}_1^2+\bar{q}_1^2)(\bar{p}_2^2 +
\bar{q}_2^2)}{\sqrt{a(a+4)}} + \frac{(\bar{p}_2^2 +
\bar{q}_2^2)^2}{a} \Big].
$$
This normal form is clearly integrable with quadratic first
integrals $I_j = \bar{p}_j^2 + \bar{q}_j^2, \, j=1, 2$.

Finally, introducing symplectic polar coordinates, which are
action - angle variables
$$
\bar{q}_j = \sqrt{2I_j} \cos \varphi_j, \quad \bar{p}_j =
\sqrt{2I_j} \sin \varphi_j, \quad j = 1, 2
$$
we get
\begin{equation}
\label{3.3} \overline{H} = \sqrt{a+4} I_1 + \sqrt{a} I_2 +
\frac{3b}{8} \Big[\frac{I_1 ^2}{a+4} + 4 \frac{I_1
I_2}{\sqrt{a(a+4)}} + \frac{I_2 ^2}{a} \Big].
\end{equation}
It can easily be checked that the Kolmogorov's condition is valid.

\subsection{Three particles}

First, we make use of the phonons $(\bar{q}, \bar{p})$ (as
explained in Section 2) to transform the quadratic part of the
Hamiltonian (\ref{1.2}) $n=3$ in diagonal form
\begin{align}
\label{3.4}
H = & \, \frac{1}{2}(\bar{p}^2 _1 + \bar{p}^2 _2 + \bar{p}^2 _3)+\frac{a+3}{2}(\bar{q}^2 _1 + \bar{q}^2 _2) +\frac{a}{2}\bar{q}^2 _3  + \\
    & \frac{b}{18}\Big[\frac{9}{2}(\bar{q}^4 _1 + \bar{q}^4 _2) + 3\bar{q}^4 _3 + 6\sqrt{2} \bar{q}^3 _1 \bar{q}_3
    + 9 \bar{q}^2 _1 \bar{q}^2 _2 + 18(\bar{q}^2 _1 + \bar{q}^2 _2)\bar{q}^2 _3 - 18 \sqrt{2} \bar{q}_1 \bar{q}^2 _2 \bar{q}_3  \Big]. \nonumber
\end{align}
Further, the scaling
\begin{eqnarray*}
\bar{q}_{1,2} & \to  \frac{1}{\sqrt[4]{a+3}} \bar{q}_{1,2} , \quad \bar{p}_{1,2} &  \to  \sqrt[4]{a+3} \bar{p}_{1,2} , \\
\bar{q}_3 & \to  \frac{1}{\sqrt[4]{a}} \bar{q}_3,   \quad
\bar{p}_3 &  \to  \sqrt[4]{a} \bar{p}_3
\end{eqnarray*}
results in
\begin{align}
H = & \frac{\sqrt{a+3}}{2} \left(\bar{p}^2 _1 + \bar{q}^2 _1 + \bar{p}^2 _2 + \bar{q}^2 _2 \right) +\frac{\sqrt{a}}{2}(\bar{p}^2 _3 + \bar{q}^2 _3 ) + \nonumber\\
    & \frac{b}{18} \Big[\frac{9(\bar{q}^4 _1 + \bar{q}^4 _2)}{2(a+3)} + \frac{3\bar{q}^4 _3}{a} + \frac{6\sqrt{2}\bar{q}^3 _1 \bar{q}_3}{\sqrt[4]{a(a+3)^3}}
    + \frac{9 \bar{q}^2 _1 \bar{q}^2 _2}{a+3} + \frac{18(\bar{q}^2 _1 + \bar{q}^2 _2)\bar{q}^2 _3}{\sqrt{a(a+3)}}
    - \frac{18 \sqrt{2} \bar{q}_1 \bar{q}^2 _2 \bar{q}_3}{\sqrt[4]{a(a+3)^3}}  \Big]. \nonumber
\end{align}
Passing to the variables $(z_j, w_j), \, j=1,2,3$ (\ref{3.2}) we
notice that there is an internal resonance between the frequencies
$\Omega_1$ and $\Omega_2$. Therefore, the generators of the normal
form are
$$
z_jw_j, \, j=1,2,3 \quad \mbox{and} \quad z_1 w_2, \, z_2 w_1 .
$$
After removing the non-resonant terms, the normal form of the
(\ref{3.4}) up to order four $\overline{H} = H_2 + \overline{H}_4$ is
\begin{align}
\overline{H} = & \frac{\sqrt{a+3}}{2} \left(z_1 w_1 +z_2 w_2 \right) + \frac{\sqrt{a}}{2} z_3 w_3 + \nonumber\\
    & \frac{b}{18} \Big[\frac{27(z_1 w_1 + z_2 w_2)^2}{16(a+3)} + \frac{9(z_3 w_3)^2}{8a} +
    \frac{9(z_1 w_1 + z_2 w_2)z_3 w_3}{2 \sqrt{a(a+3)}} + \frac{9(z_1 w_2 - z_2 w_1)^2}{16(a+3)} \Big], \nonumber
\end{align}
or, returning to $(\bar{q}, \bar{p})$ we get
\begin{align}
\overline{H} =
 & \frac{\sqrt{a+3}}{2} \left(\bar{p}^2 _1 + \bar{q}^2 _1 + \bar{p}^2 _2 + \bar{q}^2 _2 \right) +\frac{\sqrt{a}}{2}(\bar{p}^2 _3 + \bar{q}^2 _3 ) + \nonumber\\
 & \frac{b}{2} \Big[\frac{3(\bar{p}^2 _1 + \bar{q}^2 _1 + \bar{p}^2 _2 + \bar{q}^2 _2)^2}{16(a+3)} + \frac{(\bar{p}^2 _3 + \bar{q}^2 _3 )^2}{8 a}
+ \frac{(\bar{p}^2 _1 + \bar{q}^2 _1 + \bar{p}^2 _2 + \bar{q}^2
_2)(\bar{p}^2 _3 + \bar{q}^2 _3 )}{2\sqrt{a(a+3)}}
  - \frac{(\bar{p}_1 \bar{q}_2 - \bar{q}_1 \bar{p}_2)^2}{4(a+3)} \Big]. \nonumber
\end{align}
This normal form is integrable with the following quadratic first
integrals
\begin{equation}
\label{3.5} F_1:= \bar{p}^2 _1 + \bar{q}^2 _1 + \bar{p}^2 _2 +
\bar{q}^2 _2, \quad G_1:= \bar{p}_1 \bar{q}_2 - \bar{q}_1
\bar{p}_2, \quad I_3 := \bar{p}^2 _3 + \bar{q}^2 _3.
\end{equation}
In order to introduce action-angle variables, we need to find the
set of regular values of the energy momentum map
$$
EM : (\bar{q}, \bar{p}) \to (F_1, G_1, I_3).
$$
In fact, this is  done in \cite{Rink2}. Denote by $U_r = \{ (F_1,
G_1, I_3)\in \mathbb{R}^3, \, F_1 > 0 , | G_1 | < F_1, \, I_3 >0
\}$. Then for all $(F_1, G_1, I_3) \in U_r$ the level sets of
$EM^{-1} (F_1, G_1, I_3)$ are diffeomorphic to 3-tori.

Let $\mathrm{arg}: \mathbb{R}^2 \setminus \{(0, 0)\} \to
\mathbb{R}/2 \pi \mathbb{Z}$ be the argument function
$\mathrm{arg} (r \cos \Phi, r \sin \Phi) \to \Phi$. Define the
following set of variables $ (F_1, G_1, I_3, \phi_1, \psi_1,
\varphi_3)$  $F_1, G_1, I_3$ as above and
\begin{align}
\label{3.6} \phi_1 := & \, \frac{1}{2} \mathrm{arg}(-\bar{p}_2 -
\bar{q}_1, \bar{p}_1 - \bar{q}_2) +
\frac{1}{2} \mathrm{arg}(\bar{p}_2 - \bar{q}_1, \bar{p}_1 + \bar{q}_2), \nonumber \\
\psi_1 := & \, \frac{1}{2} \mathrm{arg}(-\bar{p}_2 - \bar{q}_1,
\bar{p}_1 - \bar{q}_2) -
\frac{1}{2} \mathrm{arg}(\bar{p}_2 - \bar{q}_1, \bar{p}_1 + \bar{q}_2),   \\
 \varphi_3 : = & \, \arctan \bar{p}_3/ \bar{q}_3 . \nonumber
\end{align}
Using the formula $d \mathrm{arg} (x, y) = \frac{x d y - y dx}{x^2
+ y^2}$, one can verify that $(F_1, G_1, I_3, \phi_1, \psi_1,
\varphi_3)$ are indeed canonical coordinates $\sum d \bar{p}_j
\wedge \bar{q}_j = d F_1 \wedge d \phi_1 + d G_1 \wedge d \psi_1 +
d I_3 \wedge d \varphi_3$.

Then the truncated up to order 4 normal form as a function of
actions is
\begin{equation}
\label{3.7} \overline{H} = \frac{\sqrt{a+3}}{2} F_1 +
\frac{\sqrt{a}}{2} I_3 + \frac{b}{2} \Big[\frac{3 F_1 ^2}{16(a+3)}
+ \frac{F_1 I_3}{2\sqrt{a(a+3)}} - \frac{G_1 ^2}{4(a+3)} +
\frac{I_3 ^2}{8 a}\Big].
\end{equation}
Then one can easily check-up that the Kolmogorov's condition is
valid.

{\bf Remark 2.} If we set $a=1$, we get
\begin{align}
\overline{H} =
 & \frac{2}{2} \left(\bar{p}^2 _1 + \bar{q}^2 _1 + \bar{p}^2 _2 + \bar{q}^2 _2 \right) +\frac{1}{2}(\bar{p}^2 _3 + \bar{q}^2 _3 ) + \nonumber\\
 & \frac{b}{2} \Big[\frac{3(\bar{p}^2 _1 + \bar{q}^2 _1 + \bar{p}^2 _2 + \bar{q}^2 _2)^2}{64} + \frac{(\bar{p}^2 _3 + \bar{q}^2 _3 )^2}{8}
+ \frac{(\bar{p}^2 _1 + \bar{q}^2 _1 + \bar{p}^2 _2 + \bar{q}^2
_2)(\bar{p}^2 _3 + \bar{q}^2 _3 )}{4}
  - \frac{(\bar{p}_1 \bar{q}_2 - \bar{q}_1 \bar{p}_2)^2}{16} \Big], \nonumber
\end{align}
which is an example of an integrable KAM non-degenerate normal
form of 2 : 2 : 1 (or 1 : 2 : 2) Hamiltonian resonance (see e.g.
\cite{V2}).

\subsection{Four particles}

It turns out that the normal form of the periodic KG lattice in
the case of four particles is surprisingly simple, no matter of
the internal resonance.

After transforming the quadratic part in diagonal form and scaling
\begin{eqnarray*}
\bar{q}_{1,3} & \to  \frac{1}{\sqrt[4]{a+2}} \bar{q}_{1,3} , \quad \bar{p}_{1,3} &  \to  \sqrt[4]{a+2} \bar{p}_{1,3} , \\
\bar{q}_{2} & \to  \frac{1}{\sqrt[4]{a+4}} \bar{q}_{2} , \quad \bar{p}_{2} &  \to  \sqrt[4]{a+4} \bar{p}_{2} , \\
\bar{q}_4 & \to  \frac{1}{\sqrt[4]{a}} \bar{q}_4, \, \, \, \,
\quad \bar{p}_4 &  \to  \sqrt[4]{a} \bar{p}_4
\end{eqnarray*}
the Hamiltonian (\ref{1.2}) $n=4$ takes the form
\begin{align}
\label{3.11} H = & \frac{\sqrt{a+2}}{2} (\bar{p}^2 _1 + \bar{q}^2
_1) + \frac{\sqrt{a+4}}{2} (\bar{p}^2 _2 + \bar{q}^2 _2) +
\frac{\sqrt{a+2}}{2} (\bar{p}^2 _3 + \bar{q}^2 _3) + \frac{\sqrt{a}}{2}(\bar{p}^2 _4 + \bar{q}^2 _4 ) + \nonumber\\
    & \frac{b}{8} \Big[\frac{2(\bar{q}^4 _1 + \bar{q}^4 _3)}{a+2)} + \frac{\bar{q}^4 _2}{a+4} + \frac{\bar{q}^4 _4}{a} +
    \frac{12 \bar{q}_2 \bar{q}^2 _3 \bar{q}_4}{\sqrt[4]{a(a+4)(a+2)^2}} - \frac{12 \bar{q}^2 _1 \bar{q}_2 \bar{q}_4}{\sqrt[4]{a(a+4)(a+2)^2}} \\
    & +\frac{6 \bar{q}^2 _2 (\bar{q}^2 _1 + \bar{q}^2 _3)}{\sqrt{(a+2)(a+4)}}
    +\frac{6 \bar{q}^2 _4 (\bar{q}^2 _1 + \bar{q}^2 _3)}{\sqrt{a(a+2)}} +\frac{6 \bar{q}^2 _2 \bar{q}^2 _4}{\sqrt{a(a+4)}}  \Big]. \nonumber
\end{align}
There is an internal resonance between the frequencies $\Omega_1$
and  $\Omega_3$. In variables $(z_j , w_j), \, j=1,2,3,4$, the
generators of the normal form are
$$
z_j w_j, \quad  \mbox{and} \quad  z_1 w_3, z_3 w_1 .
$$
However, as it is seen from (\ref{3.11}) the variables $\bar{q}_1$
and $\bar{q}_3$ do not couple, so the last two generators do not
appear in the normal form as if there are no resonances. After
removing the non-resonant terms, the normal form of (\ref{3.11})
up to order four $\overline{H} = H_2 + \overline{H}_4$ reads
\begin{align}
\overline{H} =
& \frac{\sqrt{a+2}}{2} z_1 w_1 + \frac{\sqrt{a+4}}{2} z_2 w_2 + \frac{\sqrt{a+2}}{2} z_3 w_3 + \frac{\sqrt{a}}{2} z_4 w_4 + \nonumber\\
& \frac{b}{8} \Big[ \frac{3 (z_1 w_1)^2}{4(a+2)} + \frac{3 (z_2 w_2)^2}{8(a+4)} + \frac{3 (z_3 w_3)^2}{4(a+2)} + \frac{3 (z_4 w_4)^2}{8 a} \nonumber \\
& +\frac{3 z_2 w_2 (z_1 w_1 + z_3
w_3)}{\sqrt{2(a+2)(a+4)}}+\frac{3 z_4 w_4 (z_1 w_1 + z_3
w_3)}{\sqrt{2a(a+2)}}+\frac{3 z_2 w_2 z_4 w_4}{\sqrt{2a(a+4)}}
\Big], \nonumber
\end{align}
or, in $(\bar{q}, \bar{p})$ variables
\begin{align}
\overline{H} = & \frac{\sqrt{a+2}}{2} (\bar{p}^2 _1 + \bar{q}^2
_1) + \frac{\sqrt{a+4}}{2} (\bar{p}^2 _2 + \bar{q}^2 _2)
+ \frac{\sqrt{a+2}}{2} (\bar{p}^2 _3 + \bar{q}^2 _3) + \frac{\sqrt{a}}{2} (\bar{p}^2 _4 + \bar{q}^2 _4) + \nonumber\\
& \frac{b}{8} \Big[ \frac{3 (\bar{p}^2 _1 + \bar{q}^2
_1)^2}{4(a+2)} + \frac{3 (\bar{p}^2 _2 + \bar{q}^2 _2)^2}{8(a+4)}
+ \frac{3 (\bar{p}^2 _3 + \bar{q}^2 _3)^2}{4(a+2)} + \frac{3 (\bar{p}^2 _4 + \bar{q}^2 _4)^2}{8 a} \nonumber \\
& +\frac{3 (\bar{p}^2 _2 + \bar{q}^2 _2) (\bar{p}^2 _1 + \bar{q}^2
_1 + \bar{p}^2 _3 + \bar{q}^2 _3)}{\sqrt{2(a+2)(a+4)}} +\frac{3
(\bar{p}^2 _4 + \bar{q}^2 _4)(\bar{p}^2 _1 + \bar{q}^2 _1 +
\bar{p}^2 _3 + \bar{q}^2 _3)}{\sqrt{2a(a+2)}}+\frac{3 (\bar{p}^2
_2 + \bar{q}^2 _2)(\bar{p}^2 _4 + \bar{q}^2 _4) }{\sqrt{2a(a+4)}}
\Big]. \nonumber
\end{align}
This normal form is integrable with integrals $I_j = \bar{p}^2 _j
+ \bar{q}^2 _j, \, j=1, 2, 3, 4$ and the action-angle variables
are clear. The truncated normal form up to order 4 as a function
of the action variables is
\begin{align}
\overline{H} = \label{3.12}
& \frac{\sqrt{a+2}}{2} I_1 + \frac{\sqrt{a+4}}{2} I_2 + \frac{\sqrt{a+2}}{2} I_3 + \frac{\sqrt{a}}{2} I_4 + \\
& \frac{b}{8} \Big[ \frac{3 (I_1^2 +I_3^2)}{4(a+2)} + \frac{3
I_2^2}{8(a+4)}  + \frac{3 I_4^2}{8 a} +\frac{3 I_2 (I_1 +
I_3)}{\sqrt{2(a+2)(a+4)}} +\frac{3 I_4 (I_1 +
I_3)}{\sqrt{2a(a+2)}}+\frac{3 I_2 I_4}{\sqrt{2a(a+4)}}  \Big].
\nonumber
\end{align}
and the verification of the validity of the Kolmogorov's condition is straightforward.

\subsection{Five particles}

There is a close resemblance between the two cases of tree and
five particles.

We transform the quadratic part of the Hamiltonian
(\ref{1.2}) $n=5$ using the phonons $(\bar{q}, \bar{p})$ and
scale in the usual way.
Further, we use the notations $\Omega_j$ for short. Recall that
$$
\Omega_1 = \Omega_4 = \sqrt{a + (5-\sqrt{5})/2}, \qquad \Omega_2 = \Omega_3 = \sqrt{a + (5+\sqrt{5})/2}, \qquad \Omega_5 = \sqrt{a},
$$
that is, $\Omega$ contains two internal resonances. The generators of the normal form in $(z, w)$ variables are
$$
z_j w_j,  \, j = 1, \ldots, 5 \quad \mbox{and} \quad z_1 w_4, \,
z_4 w_1, \, z_2 w_3 , \, z_3 w_2.
$$
Then the truncated up to order 4 normal form  $\overline{H} = H_2 + \overline{H}_4$ for (\ref{1.2}) $n=5$ reads
\begin{align}
\overline{H} = & \frac{\Omega_1}{2} (z_1 w_1 + z_4 w_4) + \frac{\Omega_2}{2} (z_2 w_2 + z_3 w_3) + \frac{\Omega_5}{2} z_5 w_5    \nonumber \\
&  + \frac{b}{10} \Big[\frac{9}{16\Omega_1^2} (z_1 w_1 + z_4 w_4)^2 + \frac{9}{16\Omega_2^2} (z_2 w_2 + z_3 w_3)^2 + \frac{3}{8\Omega_5^2} (z_5 w_5)^2   \nonumber \\
&  + \frac{3}{16\Omega_1^2} (z_1 w_4 - z_4 w_1)^2 + \frac{3}{16\Omega_2^2} (z_2 w_3 - z_3 w_2)^2 + \frac{3}{2\Omega_1 \Omega_2} (z_1 w_1 + z_4 w_4) (z_2 w_2 + z_3 w_3)   \nonumber \\
&  + \frac{3}{2\Omega_1 \Omega_5} (z_1 w_1 + z_4 w_4) z_5 w_5 + \frac{3}{2\Omega_2 \Omega_5} (z_2 w_2 + z_3 w_3) z_5 w_5  \Big],
\nonumber
\end{align}
or, written in the coordinates $(\bar{q}, \bar{p})$ is
\begin{align}
\overline{H} = & \frac{\Omega_1}{2} (\bar{p}^2 _1 + \bar{q}^2 _1 + \bar{p}^2 _4 + \bar{q}^2 _4)
+ \frac{\Omega_2}{2} (\bar{p}^2 _2 + \bar{q}^2 _2 + \bar{p}^2 _3 + \bar{q}^2 _3) + \frac{\Omega_5}{2} (\bar{p}^2 _5 + \bar{q}^2 _5)   \nonumber \\
&  + \frac{b}{10} \Big[\frac{9}{16\Omega_1^2} (\bar{p}^2 _1 + \bar{q}^2_1 + \bar{p}^2 _4 + \bar{q}^2 _4)^2
+ \frac{9}{16\Omega_2^2} (\bar{p}^2 _2 + \bar{q}^2 _2 + \bar{p}^2 _3 + \bar{q}^2 _3)^2 + \frac{3}{8\Omega_5^2} (\bar{p}^2 _5 + \bar{q}^2 _5)^2   \nonumber \\
&  - \frac{3}{4\Omega_1^2} (\bar{p}_1 \bar{q}_4 - \bar{q}_1 \bar{p}_4)^2 - \frac{3}{4\Omega_2^2} (\bar{p}_2 \bar{q}_3 - \bar{q}_2 \bar{p}_3 )^2
+ \frac{3}{2\Omega_1 \Omega_2} (\bar{p}^2 _1 + \bar{q}^2 _1 + \bar{p}^2 _4 + \bar{q}^2 _4) (\bar{p}^2 _2 + \bar{q}^2 _2 + \bar{p}^2 _3 + \bar{q}^2 _3)   \nonumber \\
&  + \frac{3}{2\Omega_1 \Omega_5} (\bar{p}^2 _1 + \bar{q}^2 _1 + \bar{p}^2 _4 + \bar{q}^2 _4) (\bar{p}^2 _5 + \bar{q}^2 _5) +
\frac{3}{2\Omega_2 \Omega_5} (\bar{p}^2 _2 + \bar{q}^2 _2 + \bar{p}^2 _3 + \bar{q}^2 _3) (\bar{p}^2 _5 + \bar{q}^2 _5)  \Big]. \nonumber
\end{align}
As it is seen, the above normal form is integrable with the following first integrals
\begin{align}
\label{3.21}
F_1 : = & \bar{p}^2 _1 + \bar{q}^2 _1 + \bar{p}^2 _4 + \bar{q}^2 _4, \, F_2 := \bar{p}^2 _2 + \bar{q}^2 _2 + \bar{p}^2 _3 + \bar{q}^2 _3 \\
G_1 : = & \bar{p}_1 \bar{q}_4 - \bar{q}_1 \bar{p}_4, \, G_2 := \bar{p}_2 \bar{q}_3 - \bar{q}_2 \bar{p}_3, \, I_5 := \bar{p}^2 _5 + \bar{q}^2 _5 . \nonumber
\end{align}
In a similar way as in \cite{Rink2} and above treated case of tree particles, in the domain where the first integrals are
independent, we can introduce action-angle variables
$$
(\bar{q}, \bar{p}) \to (F_1, F_2, G_1, G_2, I_5, \phi_1, \phi_2,
\psi_1, \psi_2, \varphi_5),
$$
where $F_j, G_j, \, j=1,2$ are as above and
\begin{align}
\label{3.22}
\phi_j := & \, \frac{1}{2} \mathrm{arg}(-\bar{p}_{5-j} - \bar{q}_j, \bar{p}_j -
\bar{q}_{5-j}) +
\frac{1}{2} \mathrm{arg}(\bar{p}_{5-j} - \bar{q}_j, \bar{p}_j + \bar{q}_{5-j}), \, j=1,2 \nonumber \\
\psi_j := & \, \frac{1}{2} \mathrm{arg}(-\bar{p}_{5-j} -
\bar{q}_j, \bar{p}_j - \bar{q}_{5-j}) -
\frac{1}{2} \mathrm{arg}(\bar{p}_{5-j} - \bar{q}_j, \bar{p}_j + \bar{q}_{5-j}), \, j=1,2   \\
 \varphi_5 : = & \, \arctan \bar{p}_5/ \bar{q}_5 . \nonumber
\end{align}
As above one can verify that $(F_1, F_2, G_1, G_2, I_5, \phi_1,
\phi_2, \psi_1, \psi_2, \varphi_5) $ are canonical coordinates.
Then the truncated up to order 4 normal form as a function of the
action variables is
\begin{align}
\overline{H} & =  \frac{\Omega_1}{2} F_1 + \frac{\Omega_2}{2} F_2 + \frac{\Omega_5}{2} I_5
  + \frac{b}{10} \Big[\frac{9}{16\Omega_1^2} (F_1)^2 + \frac{9}{16\Omega_2^2} (F_2)^2 + \frac{3}{8\Omega_5^2} (I_5)^2    \\
&  - \frac{3}{4\Omega_1^2} (G_1)^2 - \frac{3}{4\Omega_2^2} (G_2)^2 +
\frac{3}{2\Omega_1 \Omega_2} F_1 F_2  + \frac{3}{2\Omega_1 \Omega_5} F_1 I_5 + \frac{3}{2\Omega_2 \Omega_5} F_2 I_5  \Big]. \nonumber
\end{align}
It is straightforward to be checked up that the Kolmogorov's condition is valid.

\subsection{Six particles}

 In the diagonalized Hamiltonian system (\ref{1.2}) $n=6$ we scale
\begin{eqnarray*}
\bar{q}_{1,5} & \to  \frac{1}{\sqrt[4]{a+1}} \bar{q}_{1,5} , \quad \bar{p}_{1,5} &  \to  \sqrt[4]{a+1} \bar{p}_{1,5} , \\
\bar{q}_{2,4} & \to  \frac{1}{\sqrt[4]{a+3}} \bar{q}_{2,4} , \quad \bar{p}_{2,4} &  \to  \sqrt[4]{a+3} \bar{p}_{2,4} , \\
\bar{q}_{3} & \to  \frac{1}{\sqrt[4]{a+4}} \bar{q}_{3} , \quad \bar{p}_{3} &  \to  \sqrt[4]{a+4} \bar{p}_{3} , \\
\bar{q}_6 & \to  \frac{1}{\sqrt[4]{a}} \bar{q}_6, \, \, \, \,
\quad \bar{p}_6 &  \to  \sqrt[4]{a} \bar{p}_6
\end{eqnarray*}
 Recall that $\Omega_1 = \Omega_5 = \sqrt{a+1}, \Omega_2 = \Omega_4 = \sqrt{a+3}, \Omega_3 = \sqrt{a+4}, \Omega_6 = \sqrt{a}
 $, i.e., in these case we again have  two independent 1 : 1 resonances.
In the variables $(z_j, w_j), \, j = 1, \ldots, 6$ the generators of
the normal form are
$$
z_j w_j, \quad \mbox{and} \quad z_1 w_5, \, w_1 z_5, \, z_2 w_4,
\, z_4 w_2 .
$$
We normalize and find that the normal form $\overline{H} = H_2 +
\overline{H}_4$ up to order 4 is
\begin{align}
\overline{H} &=
 \frac{\Omega_1}{2} (z_1 w_1 + z_5 w_5) + \frac{\Omega_2}{2} (z_2 w_2 + z_4 w_4) + \frac{\Omega_3}{2} z_3 w_3  + \frac{\Omega_6}{2} z_6 w_6    \nonumber \\
&  + \frac{b}{72} \Big[\frac{27}{8\Omega_1^2} (z_1 w_1 + z_5 w_5)^2 + \frac{27}{8\Omega_2^2} (z_2 w_2 + z_4 w_4)^2 + \frac{9}{4\Omega_3^2} (z_3 w_3)^2  + \frac{9}{4\Omega_6^2} (z_6 w_6)^2 \nonumber \\
&  + \frac{9(z_3 w_3) (z_6 w_6)}{\Omega_3 \Omega_6} + \frac{9z_3 w_3}{\Omega_1 \Omega_3}(z_1 w_1 + z_5 w_5) + \frac{9z_3 w_3}{\Omega_2 \Omega_3}(z_2 w_2 + z_4 w_4)   \nonumber \\
&  + \frac{9z_6 w_6}{\Omega_1 \Omega_6}(z_1 w_1 + z_5 w_5) + \frac{9z_6 w_6}{\Omega_2 \Omega_6}(z_2 w_2 + z_4 w_4) +
  \frac{9}{8}\left( \frac{z_1 w_5 - w_1 z_5}{\Omega_1} - \frac{z_2 w_4 - w_2 z_4}{\Omega_2} \right)^2 \nonumber \\
& + \frac{9}{2\Omega_1 \Omega_2} \big(z_1 w_1 z_4 w_4 + z_2 w_2 z_5 w_5 + 3z_1 w_1 z_2 w_2 + 3z_4 w_4 z_5 w_5 \nonumber \\
& - (z_1w_5 + w_1 z_5)(z_2w_4 + w_2z_4) + \frac{1}{2} (z_1w_5 -
w_1 z_5)(z_2 w_4 - w_2 z_4) \big) \Big]. \nonumber
\end{align}
Before returning to the variables $(\bar{q}, \bar{p})$ we denote
\begin{equation}
\label{3.31}
 F_1 : =  \bar{p}^2 _1 + \bar{q}^2 _1 + \bar{p}^2 _5 + \bar{q}^2 _5, \quad F_2 := \bar{p}^2 _2 + \bar{q}^2 _2 + \bar{p}^2 _4 + \bar{q}^2 _4, \quad
I_3 := \bar{p}^2 _3 + \bar{q}^2 _3, \quad I_6 := \bar{p}^2 _6 + \bar{q}^2 _6 .
\end{equation}
Then the truncated normal form $\overline{H} = H_2 +
\overline{H}_4$ becomes
\begin{align}
\label{3.32}
\overline{H} &=
 \frac{\Omega_1}{2} F_1 + \frac{\Omega_2}{2} F_2 + \frac{\Omega_3}{2} I_3  + \frac{\Omega_6}{2} I_6 +
 \frac{b}{8} \Big[\frac{3 F_1^2}{8\Omega_1^2}  + \frac{3 F_2^2}{8\Omega_2^2} + \frac{I_3^2}{4\Omega_3^2}  + \frac{I_6^2}{4\Omega_6^2}  + \frac{F_1 F_2}{2\Omega_1 \Omega_2} + \frac{I_3 I_6}{\Omega_3 \Omega_6}   \nonumber \\
&  + \frac{I_3 F_1}{\Omega_1 \Omega_3}  + \frac{I_3 F_2}{\Omega_2 \Omega_3} + \frac{I_6 F_1}{\Omega_1 \Omega_6} + \frac{I_6 F_2}{\Omega_2 \Omega_6} - \frac{1}{2} \left( \frac{\bar{p}_1 \bar{q}_5 - \bar{q}_1 \bar{p}_5}{\Omega_1} - \frac{\bar{p}_2 \bar{q}_4 - \bar{q}_2 \bar{p}_4}{\Omega_3} \right)^2\nonumber \\
&  + \frac{1}{\Omega_1 \Omega_2} \Big((\bar{p}^2 _1 + \bar{q}^2 _1)(\bar{p}^2 _2 + \bar{q}^2 _2 )   + (\bar{p}^2 _4 + \bar{q}^2 _4)(\bar{p}^2 _5 + \bar{q}^2 _5))  \\
&  - 2(\bar{q}_1 \bar{q}_5 + \bar{p}_1 \bar{p}_5)(\bar{q}_2 \bar{q}_4 + \bar{p}_2 \bar{p}_4) -  (\bar{p}_1 \bar{q}_5 -
\bar{q}_1 \bar{p}_5)(\bar{p}_2 \bar{q}_4 - \bar{q}_2 \bar{p}_4)\Big) \Big]. \nonumber
\end{align}
This normal form is integrable: the independent first integrals are
$$
F_1, \,  F_2, \, I_3, \, I_6, \,
G : = (\bar{p}_1 \bar{q}_5 - \bar{q}_1 \bar{p}_5) - (\bar{p}_2 \bar{q}_4 - \bar{q}_2 \bar{p}_4) \quad \mbox{and} \quad \overline{H}_4 .
$$
However, the construction of all action-angle variables is unclear so far.

This finishes the proof of Theorem \ref{th2}.

$\hfill \blacksquare$

\section{Concluding remarks}

This paper presents partial results on integrability of normal forms of the periodic KG lattices.

We study the normal forms because the original systems are non-integrable. This is proven rigorously in the case
of two degrees of freedom (Theorem \ref{th1}) and that is one of the differences with the FPU chain.
One can carry out the non-integrability proof for $n=3$ in the same line, but with more efforts. There is a
technical difficulty to carry through that proof in the higher dimensions, however.
The variational equation (VE) does not split nicely and one needs either a tool to deal with higher dimensional (NVE)
or another particular solution with the (VE) along it suitable enough.
Nevertheless, we claim that the periodic KG lattice is non-integrable for all $n \geq 2$.

The calculation of the normal forms goes in the standard way, because we treat only low dimensions. It is also facilitated by
the assumption an $a$. It is easy to see that there are plenty of resonances when $a \in \mathbb{Q}$.
The result in Theorem \ref{th2} allows us to view the periodic KG Hamiltonian (\ref{1.2}) as a perturbation of a non-degenerate
Liouville integrable Hamiltonian, namely the truncated up to order four  Birkhoff-Gustavson normal form.
One can also verify  the other KAM condition, known as Arnold-Moser's condition \cite{A1}.

A closer look at the resonant relations in the above cases shows up that there are no third order resonant terms.
This suggests that we can also study a potential $V (x)$ of the form
\begin{equation}
\label{4.1}
V (x) = \frac{a}{2} x^2 + \frac{\gamma}{3} x^3 + \frac{b}{2} x^4 .
\end{equation}
The normalization for the periodic KG lattices in the cases up to six particles
results in $\overline{H}_3 = 0$. Therefore, Theorem \ref{th2} remains valid also
for the potentials (\ref{4.1}).

Finally, we do not address here the symmetric invariant manifolds
in the KG lattices because they can be retrieved from \cite{Rink3}.

In any case, the results of this paper serve to understand the
lattices with many particles. We intend to study them with the
approach of Rink \cite{Rink2}, based on using the symmetry
properties, which we also enjoy here, to construct suitable normal
forms.

\vspace{3ex}

{\bf Acknowledgements.} The author acknowledges funding from grant
DN 02-5 of Bulgarian Fond "Scientific Research".


\appendix

\section{Non-integrability of periodic KG lattice with $n=2$}

In this appendix we give the proof of Theorem \ref{th1}. It is
based on Ziglin-Morales-Ruiz-Ramis theory. The main result of this
theory merely says that if a Hamiltonian system is completely
integrable then the identity component of the Galois group of the
variational equation along certain particular solution is abelian.

In the applications if one finds out that the identity component
of the Galois group is non-commutative, then this implies
non-integrability. However, if this component turns out to be
abelian, one needs additional steps to prove non-integrability as
it is carried out below.

The necessary facts and results about differential Galois theory
and its relations with the integrability of Hamiltonian systems,
enough for our purposes, are written succinctly in \cite{CG}
section 2, and we do not repeat them here. We refer the reader to
\cite{M,MRS1,MR2,SvP}  for a more detailed exposition.

The proof goes in the following lines. We obtain a particular
solution and write the variational equation along this solution.
It appears that the identity component of its differential Galois
group is abelian. In order to obtain an obstacle to the
integrability, we study the higher variational equations. Their
differential Galois groups are in principle solvable. One possible
way to show that some of them is not abelian is to find a
logarithmic term in the corresponding solution (see
\cite{MRS1,MR2}). We obtain such a logarithmic term in the
solution of the second variational equation when $b \neq 0$. Then
the non-integrability of the Hamiltonian system follows.

{\bf Proof.} Suppose $b \neq 0$. First we bring the  the quadratic
part into  diagonal form. For this purpose we perform a symplectic
change of variables in (\ref{a1}) $q = Y \bar{q}, p = Y \bar{p}$,
$q = (q_1, q_2), p = (p_1, p_2)$ with $Y = \frac{1}{\sqrt{2}}
\begin{pmatrix}
1 & 1 \\
-1 & 1
\end{pmatrix}
$ (see section 2). In the new coordinates the Hamiltonian reads
(we skip the bars for simplicity here)
\begin{equation}
\label{a2} H = \frac{1}{2}\left(p_1^2 + p_2^2\right)  +
\frac{1}{2}\big[(4+a) q_1^2 + a q_2^2 \big] +
\frac{b}{4}\left(q_1^4 + 6 q_1^2 q_2^2 + q_2^4\right).
\end{equation}
{\bf Remark 3.} The Hamiltonian (\ref{a2}) is of the form
$$
H = \frac{1}{2}\left(p_1^2 + p_2^2\right)  +
\frac{1}{2}\big[\alpha_1 q_1^2 + \alpha_2 q_2^2 \big] +
\frac{b}{4}\left(q_1^4 + 6 q_1^2 q_2^2 + q_2^4\right),
$$
which is known to be integrable when $\alpha_1 = \alpha_2$ (see
for instance \cite{Kasp} and the references therein). However,
this is not the case here. We didn't succeed in finding a
non-integrability proof for the case $\alpha_1 \neq \alpha_2$,
that is why we present it here.

It is also assumed that throughout this appendix all variables are
complex: $t \in \mathbb{C}, q_j \in \mathbb{C}, p_j \in
\mathbb{C}, j = 1, 2$. The following Proposition is immediate.
\begin{prop}
\label{prop1} The Hamiltonian system corresponding to (\ref{a2})
admits a particular solution
\begin{equation}
\label{a3} q_1^0 (t) = sn (\sqrt{a+4 + b/2} t, k), \quad p_1^0 (t)
= \frac{d}{d t} q_1^0 (t), \quad q_2^0 (t) = p_2^0 (t) = 0,
\end{equation}
where $sn$ is the Jacobi elliptic function with the module $k =
\sqrt{\frac{-b/2}{4 + a + b/2}}$.
\end{prop}
 $ \hfill \square$

It is straightforward that $T_1 = \frac{4 K}{\sqrt{a + 4 + b/2}}$
and $T_2 = \frac{2 i K'}{\sqrt{a + 4 + b/2}}$ are the periods of
(\ref{a3}). Here $K, K'$ are the complete elliptic integrals of
the first kind. In the parallelogram of the periods, the solution
(\ref{a3}) has two simple poles
\begin{equation}
\label{a4} t_1 = \frac{iK'}{\sqrt{a + 4 + b/2}}, \quad t_2 =
\frac{2 K + i K'}{\sqrt{a + 4 + b/2}} .
\end{equation}

Denoting by $\xi^{(1)}_j = d q_j, \eta^{(1)}_j = d p_j, j=1,2$ the
variational equations (VE) (written as second order equations) are
\begin{eqnarray}
\label{a5}
\ddot{\xi}^{(1)} _1 & = & - \big[(4+a) + 3 b(q_1^0 (t))^2 \big] \xi^{(1)} _1, \\
\label{a6} \ddot{\xi}^{(1)} _2 & = & - \big[ a  + 3 b(q_1^0 (t))^2
\big] \xi^{(1)}_2.
\end{eqnarray}
Since $d H = p_1^0 (t) \eta^{(1)} _1 + (4+a) q_1^0 (t) \xi^{(1)}
_1 + b (q_1^0 (t))^3 \xi^{(1)} _1 $ does not depend on $(\xi^{(1)}
_2, \eta^{(1)}_2)$, then (\ref{a6}) stands for normal variational
equation (NVE), to be more specific
\begin{equation}
\label{a7} \ddot{\xi}^{(1)} _2  + \big[ a  + 3 b sn^2
(\sqrt{4+a+b/2}t, k) \big] \xi^{(1)} _2 = 0.
\end{equation}
This equation has regular singularities at $t_{1,2}$, that is, it
is a Fuchsian one.

From the expansion of the $sn$ in the neighborhood of the pole
$t_1$, we have
\begin{equation}
\label{a8} q_1^0 (t)  = \frac{1}{\sqrt{-b/2}}\Big[\frac{1}{t-t_1}
+ \frac{4+a}{6}(t-t_1) + c(t-t_1)^3 + d (t-t_1)^5 + \ldots) \Big],
\end{equation}
where $c$ is an arbitrary constant and
\begin{equation}
\label{a9} d = \frac{1}{14} \Big[ \frac{16}{27} + 4 c + \frac{4
a}{9} + ac + \frac{a^2}{9} + \frac{a^3}{108}\Big].
\end{equation}
It is not difficult to see from (\ref{a7}) and (\ref{a8}) that the
indicial equation \cite{WW} at $t=t_1$
$$ r (r-1)  - 6 =0 $$
has roots $r_{1,2} = -2, 3$ (compare with the expansions below).
Therefore, the monodromy around $t_1$ is trivial and can not serve
as an obstacle to integrability \cite{Zig} (similarly for the
monodromy around $t_2$).

\vspace{3ex}

In fact, we can say more about the identity component of the
Galois group of (VE) (\ref{a5}, \ref{a6}). Notice that each of the
variational equations (VE) is a Lam\'{e} equation in Jacobi form.

\begin{prop}
\label{prop2} The identity component of the differential Galois
group of (VE) (\ref{a5}), (\ref{a6}) is abelian.
\end{prop}
{\bf Proof.} The analysis is facilitated by the fact that (VE) is
slitted into two second order differential  equations with the
coefficients in the field of elliptic functions.

Let us start with the first equation (\ref{a5}) (the consideration
of the second equation is similar). It is straightforward that one
solution of (\ref{a5}) is $\xi^{(1)} _{1,1} = \dot{q}^0 _1 (t)$.
This solution belongs to the field of the coefficients. The other
linearly independent solution is
$$
\xi^{(1)} _{1,2} = \xi^{(1)} _{1,1} \int \frac{d
t}{\left(\xi^{(1)} _{1,1}\right)^2} ,
$$
which does not belong in general to the coefficient field. Then
the identity component of its Galois group is isomorphic to $
\begin{pmatrix}
1 & 0 \\
\nu_1 & 1
\end{pmatrix}
$. Therefore, the identity component of the Galois group of (VE)
is isomorphic to
$$
  \begin{pmatrix}

 \begin{matrix}
 1 & 0\\
 \nu_1 & 1
 \end{matrix} &
 \begin{matrix}
 0 & 0 \\ 0 & 0 \end{matrix}  \\

 \begin{matrix}
 0 & 0\\ 0 & 0 \end{matrix} &
 \begin{matrix}1 & 0 \\
 \nu_{2} & 1
 \end{matrix}
  \end{pmatrix}, \quad \nu_1, \nu_2 \in \mathbb{C}
  $$
  and it is clearly abelian. $ \hfill \square$

\vspace{3ex}

In what follows we need the expansions around $t_1$ of the
fundamental systems of solutions with unit Wronskians for
(\ref{a5}) and (\ref{a6}). For the expansions of (\ref{a5}) we get
\begin{eqnarray}
\label{a10} \xi^{(1)} _{1,1} & = & \frac{1}{(t-t_1)^2} -
\frac{4+a}{6} - 3c(t-t_1)^2 -
\left(\frac{8+6a}{27}  + 2c + \frac{ac}{2} + \frac{a^2}{18} + \frac{a^3}{216} \right)(t-t_1)^4 + \ldots  \nonumber \\
\xi^{(1)} _{1,2} & = & \frac{1}{5}(t-t_1)^3 + \ldots
\end{eqnarray}
and similarly for (\ref{a5})
\begin{eqnarray}
\label{a11} \xi^{(1)} _{2,1} & = & \frac{1}{(t-t_1)^2} -
\frac{8+a}{6} - \frac{a+6-9c}{3}(t-t_1)^2 +
 \left(\frac{56+11a}{27}  - \frac{91c+9ac}{18} - \frac{a^2}{54} - \frac{a^3}{216} \right)(t-t_1)^4 + \ldots   \nonumber\\
\xi^{(1)} _{2,2} & = & \frac{1}{5}(t-t_1)^3 + \ldots .
\end{eqnarray}
One can see that these expansions are in fact convergent since
$t_1$ is a regular singular point (cf. \cite{WW}). Hence, the
fundamental matrix $X (t)$ of (VE) is
\begin{equation}
\label{a12} X (t) =
\begin{pmatrix}

\begin{matrix}
\xi_{1,1} ^{(1)} & \xi_{1,2} ^{(1)} \\
\dot{\xi}_{1,1} ^{(1)} & \dot{\xi}_{1,2} ^{(1)}
\end{matrix}
&
\begin{matrix}
0 & 0 \\
0 & 0
\end{matrix}  \\

 \begin{matrix}
 0 & 0\\
 0 & 0
 \end{matrix}
 &
 \begin{matrix}
 \xi_{2,1} ^{(1)} & \xi_{2,2} ^{(1)} \\
 \dot{\xi}_{2,1} ^{(1)} & \dot{\xi}_{2,2} ^{(1)}
 \end{matrix}

  \end{pmatrix},
 \quad
X^{-1} (t) =
\begin{pmatrix}

 \begin{matrix}
 \dot{\xi}_{1,2} ^{(1)} & -\xi_{1,2} ^{(1)}\\
 -\dot{\xi}_{1,1} ^{(1)} & \xi_{1,1} ^{(1)}
 \end{matrix}
 &
 \begin{matrix}
 0 & 0 \\
 0 & 0
 \end{matrix}  \\

 \begin{matrix}
 0 & 0\\
 0 & 0
 \end{matrix}
 &
 \begin{matrix}
 \dot{\xi}_{2,2} ^{(1)} & -\xi_{2,2} ^{(1)}\\
 -\dot{\xi}_{2,1} ^{(1)} & \xi_{2,1} ^{(1)}
 \end{matrix}

  \end{pmatrix}.
\end{equation}

Now, let us consider the higher variational equations along the
particular solution (\ref{a3}). We put
\begin{eqnarray}
\label{a13}
q_1 & = & q^0 _1 (t) + \varepsilon \xi^{(1)} _1 + \varepsilon^2 \xi^{(2)} _1 + \varepsilon^3 \xi^{(3)} _1 + \ldots, \quad p_1 = \dot{q}_1 ,\nonumber \\
q_2 & = & 0 + \varepsilon \xi^{(1)} _2 + \varepsilon^2 \xi^{(2)}
_2 + \varepsilon^3 \xi^{(3)} _2 + \ldots, \quad \quad \, \, \, p_2
= \dot{q}_2 ,
\end{eqnarray}
where $\varepsilon$ is a formal parameter and substitute these
expressions into the Hamiltonian system governed by (\ref{a2}).
Comparing the terms with the same order in $\varepsilon$ we obtain
consequently the variational equations up to any order.

The first variational equation is, of course, (\ref{a5}),
(\ref{a6}). For the second variational equation we have
\begin{eqnarray}
\label{a14}
\dot{\xi}^{(2)} _1  & = & \eta^{(2)} _1 , \quad \dot{\eta}^{(2)} _1 = - [4 + a + 3 b (q^0 _1 (t))^2 ] \xi^{(2)} _1 + K_1 , \nonumber \\
\dot{\xi}^{(2)} _2  & = & \eta^{(2)} _2 , \quad \dot{\eta}^{(2)}
_2 = - [a + 3 b (q^0 _1 (t))^2 ] \xi^{(2)} _2 + K_2 ,
\end{eqnarray}
where
\begin{eqnarray}
\label{a15}
K_1 & = & - 3 b q^0 _1 (t) \big[ (\xi^{(1)}_1)^2 + (\xi^{(1)}_2)^2 \big], \nonumber \\
K_2 & = & - 6 b q^0 _1 (t) \xi^{(1)}_1 \xi^{(1)}_2 .
\end{eqnarray}
In this way we can obtain a chain of linear non-homogeneous
differential equations
\begin{equation}
\label{a16} \dot{\xi}^{(k)} = A (t) \xi^{(k)} + f_k (\xi^{(1)},
\ldots, \xi^{(k-1)}), \quad k = 1, 2, \ldots ,
\end{equation}
where $A (t)$ is the linear part of  the Hamiltonian equations
along the particular solution and $f_1=0$. The above equation is
called $k$-th variational equation $(\mathrm{VE}_k)$. If $X (t)$
is a fundamental matrix of $(\mathrm{VE}_1)$, then the solutions
of $(\mathrm{VE}_k), k > 1$  can be found by
\begin{equation}
\label{a17} \xi^{(k)} = X (t) c (t),
\end{equation}
where
\begin{equation}
\label{a18} \dot{c} = X^{-1} (t) f_k.
\end{equation}
Let us study the local solutions of $(\mathrm{VE}_2)$. In our case
$f_2 = (0, K_1, 0, K_2)^T $ and from (\ref{a12}) we get
$$
 X^{-1} (t) f_2 = \left( -\xi^{(1)}_{1,2} K_1, \xi^{(1)}_{1,1} K_1, -\xi^{(1)}_{2,2} K_2, \  K_2 \right)^T.
$$
We are looking for a component of $ X^{-1} (t) f_2 $ with a
nonzero residuum at $t=t_1$. This would imply the appearance of a
logarithmic term. Indeed, the residue at $t=t_1$ of
$\xi^{(1)}_{2,1} K_2$ with the specific representatives is
$$
\mathrm{Res}_{t=t_1} (-6 b \xi^{(1)}_{2,1} q^0 _1 (t)
\xi^{(1)}_{1,1} \xi^{(1)}_{2,1}) =
\frac{6b}{\sqrt{-b/2}}\Big[\frac{a^3}{252} + \frac{a^2}{21} +
\frac{4a}{21} + \frac{3ac}{7} + \frac{73c+16}{63} \Big].
$$
Since $c$ is an arbitrary parameter, we choose it in such a way
that the expression in the square brackets does not vanish for
$a>0$. There are many such values of $c$, say $c=1$. Recall that
by assumption $b \neq 0$. We have obtained a nonzero residuum at
$t=t_1$, which implies the appearance of a logarithmic term in the
solutions of $(\mathrm{VE}_2)$. Then its  Galois group is solvable
but not abelian. Hence, we conclude the non-integrability of the
Hamiltonian system (\ref{a1}).

$\hfill \blacksquare$

\end{document}